\documentclass[conference]{IEEEtran}
\IEEEoverridecommandlockouts
\usepackage{cite}
\usepackage{amsmath,amssymb,amsfonts}
\usepackage{algorithmic}
\usepackage{graphicx}
\usepackage{textcomp}
\usepackage{xcolor}
\usepackage{listings}
\usepackage{todonotes}
\usepackage{bm}
\DeclareMathOperator*{\argmax}{arg\,max}
\def\BibTeX{{\rm B\kern-.05em{\sc i\kern-.025em b}\kern-.08em
    T\kern-.1667em\lower.7ex\hbox{E}\kern-.125emX}}

\begin{document}

\title{Multiobjective Reinforcement Learning for Reconfigurable Adaptive Optimal Control of Manufacturing Processes
\thanks{The authors would like to thank the German Federal Ministry of Education and Research (BMBF) for funding the presented research under grant \#03FH061PX5.}
}

\IEEEoverridecommandlockouts\IEEEpubid{\makebox[\columnwidth]{978-1-5386-5925-0/18/\$31.00~\copyright~2018 IEEE \hfill} \hspace{\columnsep}\makebox[\columnwidth]{ }}

\author{
	\IEEEauthorblockN{Johannes Dornheim}
	\IEEEauthorblockA{\textit{Intelligent Systems Research Group (ISRG)} \\
	\textit{Karlsruhe University of Applied Sciences}\\
	Karlsruhe, Germany \\
	johannes.dornheim@hs-karlsruhe.de}
	\and
	\IEEEauthorblockN{Norbert Link}
	\IEEEauthorblockA{\textit{Intelligent Systems Research Group (ISRG)} \\
	\textit{Karlsruhe University of Applied Sciences}\\
	Karlsruhe, Germany \\
	norbert.link@hs-karlsruhe.de}
}
\maketitle

\begin{abstract}
In industrial applications of adaptive optimal control often multiple contrary objectives have to be considered. The relative importance (weights) of the objectives are often not known during the design of the control and can change with changing production conditions and requirements. In this work a novel model-free multiobjective reinforcement learning approach for adaptive optimal control of manufacturing processes is proposed. The approach enables sample-efficient learning in sequences of control configurations, given by particular objective weights.
\end{abstract}

\section{Introduction}

Methods for adaptive optimal control like Dynamic Programming (DP) or Reinforcement Learning (RL) are usually defined for the optimization with respect to scalar reward-functions $R$. In contrast, in multiobjective RL (MORL) methods optimizing with respect to a vector-valued reward function $\Re$, where every vector component is related to a single objective, are investigated. In this paper, the application of MORL methods to manufacturing adaptive control with sequentially changing configurations is investigated, where each configuration is defined by an individual objective-weighting. Sample efficient adaption of the control to changing weights is aimed for via information transfer between the control configurations.

In RL, for observed Markovian states $s$, the selection of actions $a$ leading to successive states $s'$ are optimized. Optimization goal is the maximization of the expected, cumulated future rewards, gained per decision made from a reward-signal $R(s, a, s')$. $Q$-function based RL, which is treated in this work, is doing so by constantly updating an expected value function $Q(s,a)$ of the future rewards while interacting with the environment under an \textit{explorative policy}. $Q$-function based RL methods can be divided into \textit{off-policy}- and \textit{on-policy}-methods, based on the way in which the $Q$-function is updated. In on-policy methods, the $Q$-function models the expected reward for the explorative policy the agent is executing. In off-policy methods, the $Q$-function models the expected future reward for the optimal policy. Well known update mechanisms, used in this work are the off-policy $Q$-learning update and the on-policy SARSA-update. In this work Artificial Neural Networks (NN) are used for $Q$-function approximation, enabling generalization and thus sample-efficient learning. The NN training-mechanism used, is derived from the incremental variant of Neural Fitted Q-Iteration (NFQ) \cite{Riedmiller2005}, which itself is a special realization of the Fitted Q-Iteration (FQI) algorithm.

MORL \cite{Liu2015} and more general multiobjective decision making \cite{Roijers2013} approaches can be divided into single-policy- and multiple-policy-approaches. While the goal of single-policy algorithms is to find the best single-policy for given objective-preferences, the goal of multi-policy algorithms is to approximate the pareto-front of the solution-space by finding a set of policies, each optimal for specific preferences. Value based single- and multiple-policy-approaches are often using a parametrized scalarization function $f$ assigning a scalar output to the reward-vector $\Re$ or to a vector of reward-estimates, where the parameters are reflecting particular weightings (configurations). The information transfer in sequences of changing configurations, investigated in this paper, is not clearly assignable to single- or multiple-policy- approaches. Unlike in single-policy MORL, the future configurations are unknown during learning and unlike in multiple-policy MORL, the goal is not to find an explicit set of pareto-dominant policies but instead to enable information transfer in scenarios with changing configurations. Related approaches, using MORL for information transfer over objective-weightings in different application scenarios and forms are \cite{Natarajan2005}, \cite{Ngai2011} and \cite{Taylor2014}. In \cite{Natarajan2005}, R-learning combined with explicitly storing policies for different weightings is used in an multiobjective manner for the solution of time-varying weighting problems. In \cite{Ngai2011}, a control algorithm for a vehicle overtaking is proposed, where expected values for seven objectives are scalarized by applying the weighted arithmetic mean, dynamically adjusted via binary weights by a planning algorithm. In \cite{Taylor2014}, a method for parallel transfer in a multi-agent and multiobjective smart-grid optimization scenario is proposed. One approach, applying a Fitted Q-Iteration approach to a MORL setting has been identified \cite{Castelletti2011}. In \cite{Castelletti2011} FQI is used for offline multiple-policy MORL based on prior sampling and a scalar $Q$-function $Q(s,a,w)$, which is generalizing over weights $w$. To the best of our knowledge, neither a previous approach is applying a FQI method to MORL in an online manner, nor an approach for sequential multiobjective information transfer in the field of process control can be found in the literature.

\section{Approach}
\label{approach}
In the multiobjective case considered, instead of learning the scalar $Q$-Function for a single reward signal, a vector-valued function $\hat{Q}(s,a) \rightarrow \mathbb{R}^I$ is learned. The $i$-th component of the functions output vector, $\hat{Q}_i$, represents the expected value for the $i$-th objective. When combined with a $Q$-Function based RL-algorithm, the change from the single-objective $Q$-Function to a multiobjective $\hat{Q}$-function affects the function update and the extraction of the agents goal- and explorative-policy from the $\hat{Q}$-function. 

For a learned $\hat{Q}$-function and given objective-weights $w$, the goal-policy is obtained by greedily acting for the expected values $\pi(s) = \argmax_a f(\hat{Q}(s,a), w)$, scalarized by a given function $f$. The explorative policy is derived from the goal-policy analogous to single-objective RL by expanding it with exploration (e.g. by acting $\epsilon$-greedy in the goal-policy).

In section \ref{eval}, off-policy and on-policy $\hat{Q}$-function updates are compared for the use in the proposed setting. The off-policy update as defined in (\ref{off_policy_update}) is applied per vector-component, where $\Re$ is the reward vector and $\gamma$ the discount-factor. The component-wise update equals the scalar $Q$-learning update rule used in single-objective RL.

\begin{equation}
\begin{split}
\hat{Q}_i(s,a) \leftarrow &(\alpha-1)\hat{Q}_i(s,a) +\\
&\alpha[\Re_i + \gamma {max}_{a_{t+1}}\hat{Q}_i(s_{t+1}, a_{t+1})]
\end{split}
\label{off_policy_update}
\end{equation}

The on-policy $\hat{Q}$-update used is defined in (\ref{on_policy_update}), where $a_{t+1}$ is the action actually executed in control-step $t+1$. Like in the off-policy case it equals the component-wise application of the scalar SARSA-update. 

\begin{equation}
\begin{split}
\hat{Q}_i(s,a) \leftarrow &(\alpha-1)\hat{Q}_i(s,a) + \\
& \alpha[\Re_i + \gamma \hat{Q}_i(s_{t+1}, a_{t+1})]
\end{split}
\label{on_policy_update}
\end{equation}

Using off-policy updates objective-wise leads to a systematic positive bias in the $\hat{Q}$-function, because it assumes a goal-policy which is greedy in the according objective and is not considering the influence of other objectives to the optimal policy. Thus, in single-policy MORL algorithms usually on-policy updates are used. However, the goal of the approach is the transfer of the learned $\hat{Q}$-function to tasks with different objective-weights and hence different optimal policies. In this case the off-policy update has the advantage of being independent of particular weightings, while the on-policy update is leading to a $\hat{Q}$-function which is dependent on the policy executed at the update-time. Due to this reason, MORL applications including an implicit or explicit transfer of the expected value function to varying objective-weights are usually based on off-policy update rules.

The multiobjective update rules and the extraction of the goal- and the explorative-policy can be combined with a wide range of $Q$-learning- or SARSA-based single-objective RL algorithms, by considering the mentioned restrictions. The proposed approach uses an adapted version of the incremental $Neural Fitted Q Iteration$ (NFQ) variant \cite{Riedmiller2005} with multiple-output neural networks (NN) and $\epsilon$-greedy exploration. In contrast to algorithms incrementally redefining the $Q$-function approximation, in NFQ the NNs are re-trained from scratch every $J$ episodes based on a sample-set $D$ in our case consisting of $(s_t,a_t, s_{t+1},\Re)$-tuples. This has two advantages in the sequential multiobjective information transfer case: (a) for both, off- and on-line updates, the effect of so called \textit{catastrophic forgetting} is avoided, which would occur, when methods without a sample-memory are used and which is caused by drastically changing configuration dependent explorative- and goal-policies. (b) The mentioned problem with on-policy updates in information transfer settings is solved by recalculating the offline update in every re-training for the current objective-weights by sampling $a_{t+1}$ on-the-fly from the current explorative-policy.

Our MORL-specific NFQ approach is adapted to the task of fixed-horizon manufacturing process control. Instead of a single NN, time-dependent $\hat{Q}$-models are used, which are trained backwards in time-steps. An in-depth description of the application-specific NFQ adoptions can be found in our previous work \cite{Dornheim2018}.

\section{Evaluation}
\label{eval}

\begin{figure}
	\includegraphics[width=0.5\textwidth]{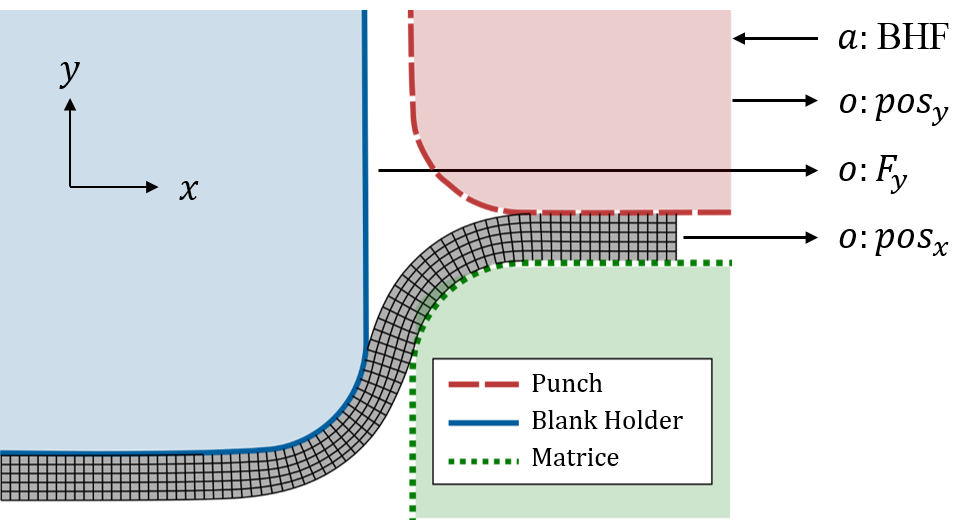}
	\caption{Rotationally symmetric deep drawing FEM simulation model with observables $o$ and control-actions $u$}
	\label{fem}
\end{figure}

The evaluation scenario for the approach proposed is the optimal control of time-varying blank holding forces (BHF) in a deep drawing process, regarding two conflicting objectives: (a) the material efficiency and (b) the stability of the resulting product. The evaluation is based on a finite element method (FEM) simulation of the deep drawing process, depicted in Fig. \ref{fem}, assuming rotational symmetry and isotropic material behavior. During a single process execution, the control agent can set BHF values from the set of $\{20kN, 40kN, ..., 140kN\}$ at five equidistant control times. Observables $o$ obtained by the agent at each control step are the actual stamp-force $F_y$, the actual blank infeed in x-direction $pos^{blank}_x$ and the actual offset of the blank-holder in y-direction $pos^{BH}_y$. Additive gausian measurement noise is applied to the observable values, where $\sigma$ is defined as $0.5\%$ of the particular value-range for $F_y$, $pos^{blank}_x$ and $0.25\%$ of the value-range for $pos^{BH}_y$). Furthermore stochastic process behavior is induced by varying friction-coefficients, randomly drawn per process-execution from a beta-distribution with $\alpha = 3$, $\beta = 15$. A detailed description of the evaluation environment can be found in \cite{Dornheim2018}.

The reward vector $\Re$ consists of two components and is only non-zero at terminal states, where $\Re$ consists of a material-efficiency- and a product-quality-reward. The first is the negated final infeed and the second is the thickness of the thinnest part of the resulting cup. Both rewards are derived from the FEM results and scaled into the range [0,10].

The scalarization function used in the evaluation scenario is the weighted harmonic mean $H(\Re, w)$. While the weighted arithmetic mean tends to prefer extreme solutions and is not able to find solutions in non-convex areas of the pareto-front (\cite{Liu2015}), the weighted harmonic mean emphasizes solutions with balanced reward-terms. Using $H(\Re,w)$ as scalarization function $f$ violates the assumption of function based RL, that  $f(\mathbb{E}[\Re])=\mathbb{E}[f(\Re)]$. For concave functions, like the weighted harmonic mean, \textit{Jensens inequality} states that $f(\mathbb{E}[X])\geq\mathbb{E}[f(X)]$ \cite{jensen1906fonctions}. To investigate how this effects the control quality in the specific evaluation scenario, the proposed MORL approach is compared to a single-policy NFQ approach which is based on a scalar $Q$-Function for the weighted arithmetic mean reward.

\begin{figure}
	\includegraphics[width=0.5\textwidth]{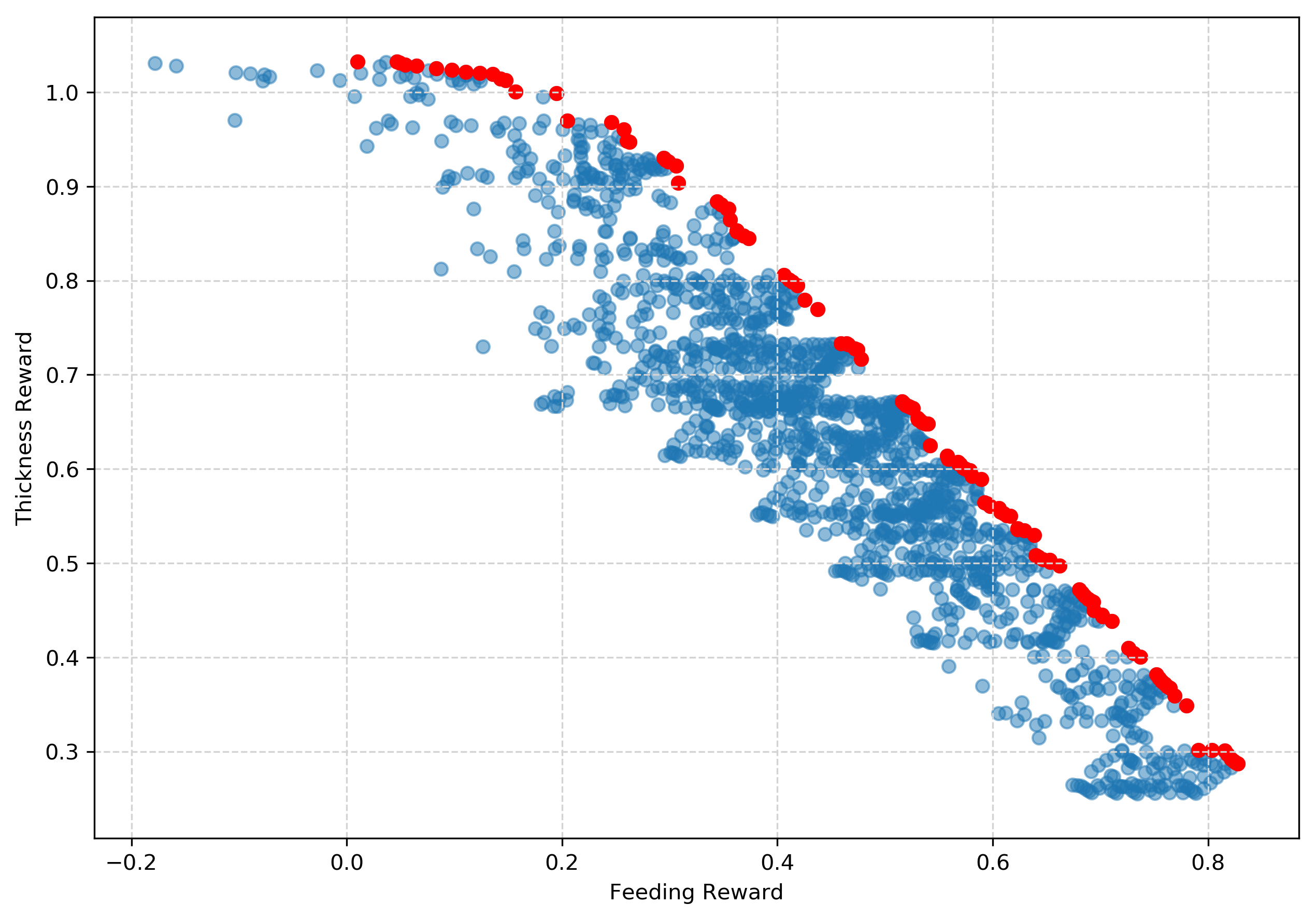}
	\caption{Solution-space with highlighted pareto-front solutions (red) for a given friction coefficient of 0.028}
	\label{pareto_front}
\end{figure}

In Fig. \ref{pareto_front} the solution-space for one specific friction coefficient, obtained by full-simulation, is plotted with pareto-front solutions marked in red.

The Neural Networks, approximating $\hat{Q}_t$ at the five time-steps, are trained by using LBFGS \cite{liu1989limited}. They consist of two hidden layers with 50 neurons each for $t\in \{2,3,4\}$, two hidden layers with 10 neurons each for $t=1$ and one hidden layer consisting of 5 neurons for $t=0$. 

Experiments where made in sequences of four tasks, representing an initial control configuration and three re-configurations. Each task is executed with different weight-values ($w_{thickness}$, $w_{infeed}$) with $w_{thickness}$ drawn randomly from the set ${1,2, ...,9}$ and $w_{infeed} = 10 - w_{thickness}$. Every task consisted of 1000 episodes, where one episode $j$ denotes one execution of the deep-drawing process. 

The experiments were executed with a learning rate $\alpha=0.7$ and an exponentially decaying exploration rate $\epsilon_i = \epsilon_0 \mathrm{e}^{-\lambda i}$ for episode $i$ with $\epsilon_0=0.1$ and $\lambda=10^{-3}$. $\hat{Q}_t$-function NNs are retrained every 50 episodes.

\section{Results}
\begin{figure*}
	\centering
	\includegraphics[width=0.9\textwidth]{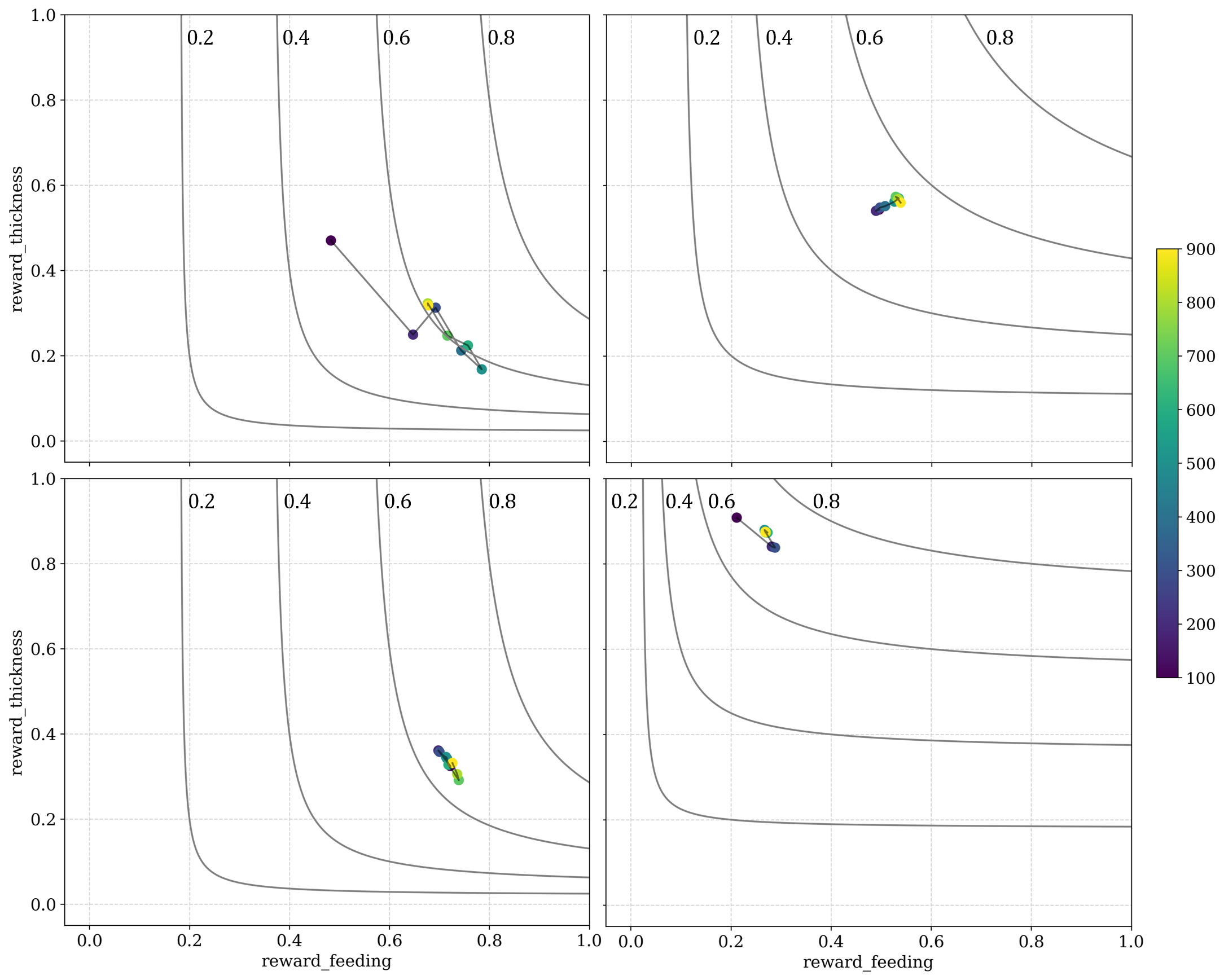}
	\caption{Moving average of the reward gained for four consecutive executions with changing term-weights of the off-policy transfer approach. Weights $w$ of the weighted harmonic mean are reflected by contour lines for $H(w)=0.2$, $H(w)=0.4$, $H(w)=0.6$ and $H(w)=0.8$. Weights of the plotted experiments are in sequential order (from top left to bottom right) $[(w_{feeding}=9.0, w_{thickness}=1.0), (w_{feeding}=5.0, w_{thickness}=5.0), (w_{feeding}=9.0, w_{thickness}=1.0), (w_{feeding}=1.0, w_{thickness}=9.0)]$}
	\label{trajectory}
\end{figure*}

\begin{figure}
	\centering
	\includegraphics[width=0.45\textwidth]{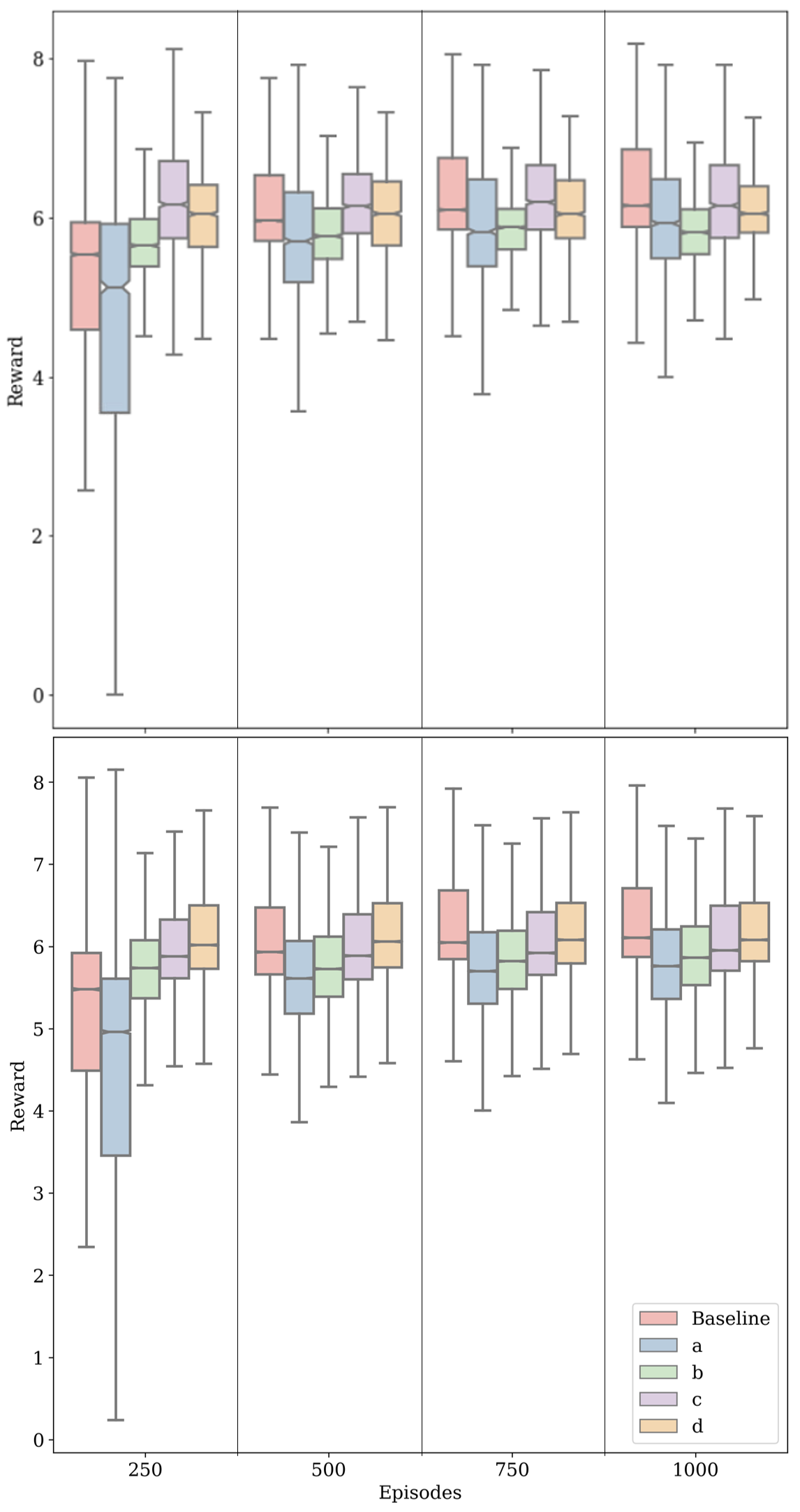}
	\caption{Aggregated reward by episodes for the on-policy update (top) and the off-policy update (bottom) for various positions in the task-sequence (a to d) in comparison to the baseline algorithm (red)}
	\label{box_vs_baseline}
\end{figure}

Results of the experiments, executed with the proposed approach, described in section \ref{eval}, are visualized in Fig. \ref{trajectory} and in Fig. \ref{box_vs_baseline}. The qualitative plot in Fig. \ref{trajectory} shows, for an exemplary single experiment, the positive effect of the information transfer regarding the optimization convergence. The quantitative plot in Fig. \ref{box_vs_baseline} visualizes data aggregated from 100 experiments with the on-policy update and 100 experiments with the off-policy update and is showing the convergence behavior over a variety of possible task sequences.

In Fig. \ref{trajectory} exemplary results for the approach with off-policy updates for a specific task-sequence are plotted. The weight-values for each configuration are depicted by contour lines of the according weighted harmonic mean function. The optimization-progress is depicted by trajectories of markers, colored by episode from dark blue to yellow. Due to the described stochastic manufacturing process and the stochastic optimization process itself, the results are very noisy and have been smoothed by applying the \textit{moving average}. For every 100 episodes a marker is reflecting the mean of gained rewards by reward term for the surrounding 200 episodes. The figure shows for the sequence that information transfer is happening, leading to faster convergence and increasing rewards over the task-sequence. 

In Fig. \ref{box_vs_baseline}, results aggregated from a series of experiments are plotted for the on- and the off-policy update. For both cases 100 experiments as described where carried out. The reward gained per episode is visualized as box-plots grouped by the position of the task-instance in the task-sequence and aggregated over the 100 experiments and over 250 episodes per box. The baseline (red box) is consisting of 100 random task-instances, independently optimized by single-policy learning with single-objective manufacturing-process NFQ, as proposed in \cite{Dornheim2018}. The plot shows a positive effect of the amount of prior knowledge to the convergence speed. While the reward reached in the first task (a, blue) is noticeably worse than the baseline (red) over all episodes, from the second task-instance (b, green) on faster convergence can be seen for both update strategies. From the third task-instance (c, violet) on, the performance after 750 episodes is in the same range as the baseline-performance for both update methods.

\begin{figure}
	\includegraphics[width=0.461\textwidth]{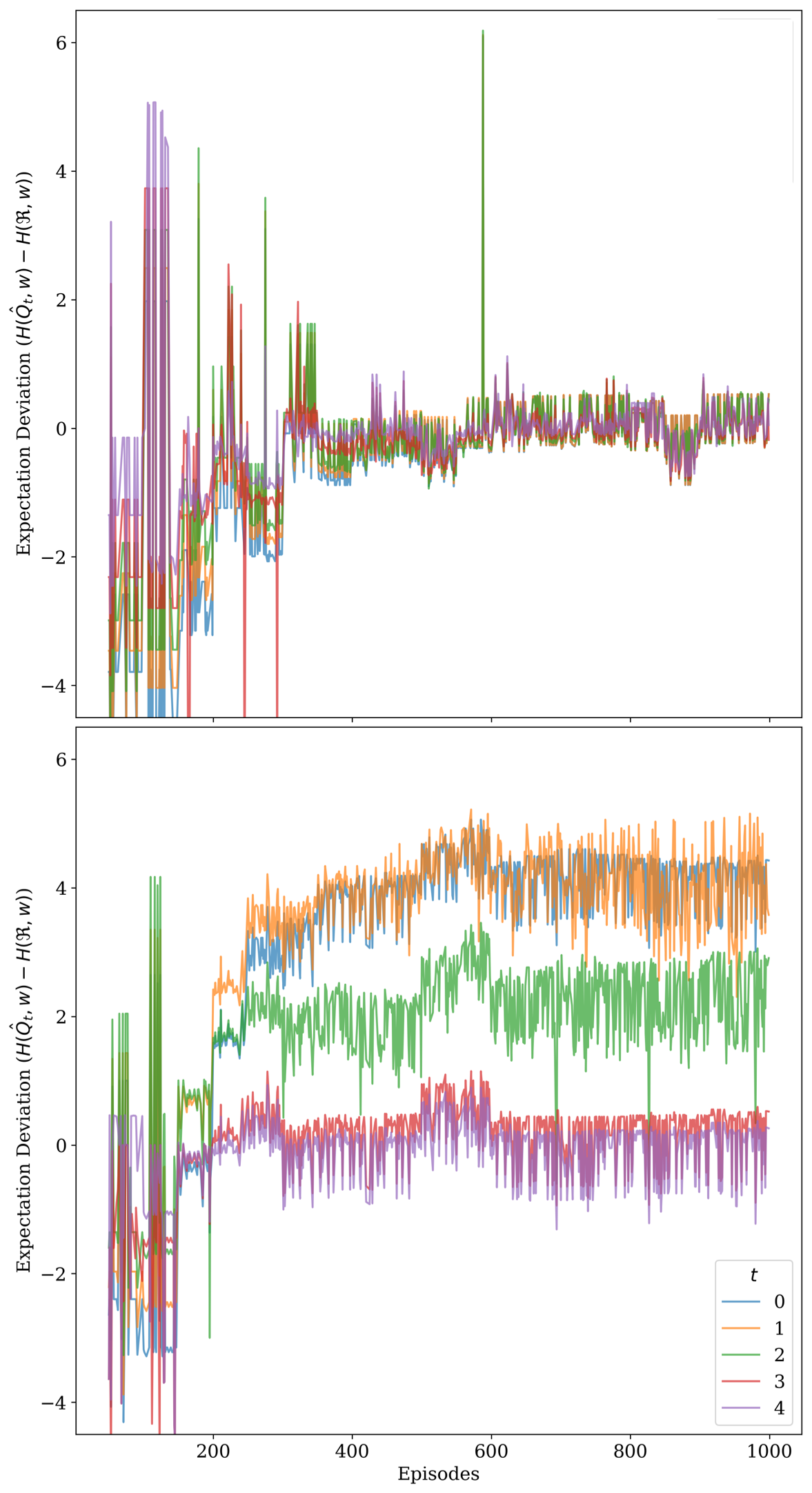}
	\centering
	\caption{Expectation deviation (expected reward - actual reward) for exemplary optimization runs by episode for expectation time-step (color) in the on-policy update (top) and the off-policy update (bottom).}
	\label{expectation_deviation}
\end{figure}

In Fig. \ref{expectation_deviation} typical examples of the deviation between the scalarized future reward ($H(\hat{Q}_t, w)$) expected in time-step $t$ and the scalarized reward ($H(\Re, w)$) observed afterwards in time-step $T$ is plotted for $t\in\{0,1,...,4\}$, for both, the on-policy update and the off-policy update. The systematic under-estimation in the first episodes in both cases is due to the initialization of the $\hat{Q}$-functions with zero-filled vectors combined with the smoothing effect of the learning rate $\alpha$. The positive-bias, described in chap. \ref{approach}, for the off-policy approach is reflected by the systematic over-estimation decreasing with the estimation-horizon $T-t$. In the on-policy case, after the initial under-estimation phase, the estimation deviation is unbiased, reflected by a mean near zero.

\section{Conclusion and Future Work}
An approach, based on NFQ for multiobjective manufacturing processes, has been proposed, enabling an information transfer between different process configurations (objective weightings). The improvement of the optimization convergence in sequences of tasks with varying objective-weights has been shown and the quantitative and qualitative difference of on-policy and off-policy updates was studied. An extension of the approach and a systematic comparison with other approaches on a wider class of Markov Decision Processes (MDP), beyond fixed-horizon MDPs with non-zero rewards only at the terminal state, is planned.

\bibliographystyle{ieeetr}
\bibliography{paper}

\begin{thebibliography}{10}

\bibitem{Riedmiller2005}
M.~Riedmiller, ``{Neural Fitted Q Iteration - First Experiences with a Data
  Efficient Neural Reinforcement Learning Method},'' in {\em ECML},
  pp.~317--328, Springer, 2005.

\bibitem{Liu2015}
C.~Liu, X.~Xu, and D.~Hu, ``{Multiobjective Reinforcement Learning: A
  Comprehensive Overview},'' {\em IEEE Transactions on Systems, Man, and Cybe},
  vol.~45, no.~3, pp.~385--398, 2015.

\bibitem{Roijers2013}
D.~M. Roijers, P.~Vamplew, S.~Whiteson, and R.~Dazeley, ``A survey of
  multi-objective sequential decision-making,'' {\em Journal of Artificial
  Intelligence Research}, vol.~48, pp.~67--113, 2013.

\bibitem{Natarajan2005}
S.~Natarajan and P.~Tadepalli, ``{Dynamic preferences in multi-criteria
  reinforcement learning},'' {\em ICML}, pp.~601--608, 2005.

\bibitem{Ngai2011}
D.~C.~K. Ngai and N.~H.~C. Yung, ``{A multiple-goal reinforcement learning
  method for complex vehicle overtaking maneuvers},'' {\em IEEE Transactions on
  Intelligent Transportation Systems}, vol.~12, no.~2, pp.~509--522, 2011.

\bibitem{Taylor2014}
A.~Taylor, I.~Dusparic, E.~Galvan-Lopez, S.~Clarke, and V.~Cahill,
  ``{Accelerating Learning in multi-objective systems through Transfer
  Learning},'' in {\em IJCNN}, 2014.

\bibitem{Castelletti2011}
A.~Castelletti, F.~Pianosi, and M.~Restelli, ``{Multi-objective fitted
  Q-iteration: Pareto frontier approximation in one single run},'' in {\em
  ICNSC 2011}, 2011.

\bibitem{Dornheim2018}
J.~Dornheim, N.~Link, and P.~Gumbsch, ``Model-free adaptive optimal control of
  sequential manufacturing processes using reinforcement learning,'' {\em arXiv
  preprint arXiv:1809.06646}, 2018.

\bibitem{jensen1906fonctions}
J.~L. W.~V. Jensen, ``Sur les fonctions convexes et les in{\'e}galit{\'e}s
  entre les valeurs moyennes,'' {\em Acta mathematica}, vol.~30, no.~1,
  pp.~175--193, 1906.

\bibitem{liu1989limited}
D.~C. Liu and J.~Nocedal, ``On the limited memory bfgs method for large scale
  optimization,'' {\em Mathematical programming}, vol.~45, no.~1-3,
  pp.~503--528, 1989.

\end{thebibliography}

\end{document}